# Masses of Four Missing Hadrons having Spin 3/2h Predicted in SU(4) by Standard Model

**Imran Khan**

*Department of Physics, University of Science and Technology,*
*Bannu 28100, Pakistan*

Email address: immarwat@yahoo.com

*Abstract*—In particle physics, study of the symmetry and its breaking play very important role in order to get useful information about the nature. The classification and arrangements of subatomic particles is also necessary to study particle physics. Particles which are building blocks of nature are quarks, gluons and leptons. Baryons and Mesons composed of quarks were arranged by Gell-Mann and Okubo in their well-known Eight-Fold way up to SU(3) symmetry. Standard model of particles is composed of these particles. Particles in SU(4) also make some beautiful patron. These make some multiplets. but all the baryons with spin $J^P = 3/2^+$ in these multiplets have not been observed till date. In this paper the SU(4) multiplets have been organized and studied in an easy and new way. As a result we obtained some clues about the masses and other characteristics of the unknown hyperons. These approximations about the characteristics of the unidentified baryons have been recorded in this article. Mass formula for the baryon SU(4) multiplets have been obtained.

**Keywords**—*Baryons, SU(3), SU(4), Hyperons, Standard Model, Mass Formula.*

## I. INTRODUCTION

Baryons are made of three quarks (*qqq*). The three flavors up *u*, down *d*, and strange *s*, imply an approximate flavor *SU*(3), which requires that baryons made of these quarks belong to the multiplets on the right side of the 'equation' $3 \otimes 3 \otimes 3 = 10_S \oplus 8_M \oplus 8_M \oplus 1_A$. Here the subscripts indicate symmetric, mixed symmetry, or anti-symmetric states under interchange of any two quarks [1]. These were classified and arranged by Gell-Mann and Okubo in De-Couplets ($J = 3/2$, $l = 0$) and Octets ($J = 1/2$, $l = 0$) with +1 parities [2–4]. The Gell-Mann / Okubo mass formula which relates the masses of members of the baryon octet is given by; [2–4]

$$2(m_N + m_\Xi) = 3m_\Lambda + m_\Sigma \qquad (1)$$

While mass formula for De-couplets consists of equal spacing between the rows. The spaces are equal to an average value − 151MeV/c².

$$M_\Delta - M_{\Sigma^*} = M_{\Sigma^*} - M_{\Xi^*} = M_{\Xi^*} - M_\Omega \qquad (2)$$

## 2. SU(4) MULTIPLETS

Now let's move towards baryon types made from the combination of four quarks, i.e. up $u$, down $d$, strange $s$ and charm $c$. These belong to $SU(4)$ multiplets. The $SU(4)$ multiplets numerology is given by

$$4 \otimes 4 \otimes 4 = 20_S \oplus 20_M \oplus 20_M \oplus 4_A \;\; [5].$$

We are interested in twenty particles having spin 3/2 and even parity +1 forming one of the $SU(4)$ multiplets. These particles are in their ground states, that is with $l = 0$. Their masses and other properties are listed in the Table 1, and shown in figure 1.

Table 1. List of the spin $J^P = 3/2^+$ particles formed from three quarks combinations ($qqq$) of four quarks $u$, $d$, $s$ and $c$ [5].

| S.# | Particle Name | Symbol | Quark contents | Rest Mass ($MeV/c^2$) |
|---|---|---|---|---|
| 1 | Delta $^{++}$ | $\Delta^{++}(1232)$ | $uuu$ | 1232±2 |
| 2 | Delta Plus | $\Delta^{+}(1232)$ | $uud$ | 1232±2 |
| 3 | Delta Zero | $\Delta^{0}(1232)$ | $udd$ | 1232±2 |
| 4 | Delta Minus | $\Delta^{-}(1232)$ | $ddd$ | 1232±2 |
| 5 | Sigma Plus | $\Sigma^{*+}(1385)$ | $uus$ | 1382.8±0.4 |
| 6 | Sigma Zero | $\Sigma^{*0}(1385)$ | $uds$ | 1383.7±1.0 |
| 7 | Sigma Minus | $\Sigma^{*-}(1385)$ | $dds$ | 1387.2±0.5 |
| 8 | charmed Sigma $^{++}$ | $\Sigma_c^{*++}(2520)$ | $uuc$ | 2517.9±0.6 |
| 9 | charmed Sigma $^{+}$ | $\Sigma_c^{*+}(2520)$ | $udc$ | 2517.5±2.3 |
| 10 | charmed Sigma Zero | $\Sigma_c^{*0}(2520)$ | $ddc$ | 2518.8±0.6 |
| 11 | Xi Zero | $\Xi^{*0}(1530)$ | $uss$ | 1531.80±0.32 |
| 12 | Xi Minus | $\Xi^{*-}(1530)$ | $dss$ | 1535.0±0.6 |
| 13 | charmed Xi $^{+}$ | $\Xi_c^{*+}(2645)$ | $usc$ | $2645.9^{+0.5}_{-0.6}$ |
| 14 | charmed Xi Zero | $\Xi_c^{*0}(2645)$ | $dsc$ | 2645.9±0.5 |
| 15 | Omega Minus | $\Omega^{-}(1672)$ | $sss$ | 1672.45±0.29 |
| 16 | charmed Omega Zero | $\Omega_c^{*0}(2770)$ | $ssc$ | 2765.9±2.0 |
| 17 | double charmed Xi $^{++}$ | $\Xi_{cc}^{*++}$ | $ucc$ | Unknown |
| 18 | double charmed Xi $^{+}$ | $\Xi_{cc}^{*+}$ | $dcc$ | Unknown |
| 19 | double charmed Omega $^{+}$ | $\Omega_{cc}^{*+}$ | $scc$ | Unknown |
| 20 | triple charmed Omega $^{++}$ | $\Omega_{ccc}^{*++}$ | $ccc$ | Unknown |

Since the mass splitting due to strangeness is typically of order 151 $MeV/c^2$ in the de-couplet, which is an effect of about 13% of the masses it contains. The splitting due to charm in the $SU(4)$ particles multiplets is expected to be larger [6]. The ratio $\frac{m_2}{m_1} = \frac{m_4}{m_3}$ indicates the direction of the increasing mass in the Strangeness, or Charm-ness planes in *SU*(3) and *SU*(4) multiplets respectively. It gives the mass symmetry breaking ratio for the multiplet [6]. In the *SU*(3) framework the Gell-Mann / Okubo relation for the $J^P = 1/2^+$ Octets, Eq. (1) and the equal spacing rule for the $J^P = 3/2^+$ de-couplets, Eq. (2) work so nicely that we cannot abandon linear mass formulae for baryons [6]. Using the above relation for baryons of $J^P = 3/2^+$ de-couplet of *SU*(3) gives a mass symmetry breaking ratio of order 1.0.

## II. METHOD

There are four particles in the Table 1, whose masses are unknown. Let's try to make some predictions about the unknown masses of the $\Xi_{cc}^{*++}, \Xi_{cc}^{*+}, \Omega_{cc}^{*+}$, and $\Omega_{ccc}^{*++}$ hyperons. The particles listed in Table 1 are grouped in pyramid shape of increasing charm number from $C = 0$ to $C = 3$ as shown in Fig. 1 [5]. Masses of the particles are expressed in round numbers. The bottom of the 'Pyramid' is the *SU*(3) de-couplet of $J^P = 3/2^+$ particles at $C = 0$. Average masses written on the right side of the Fig. 1 are discussed later in the text.

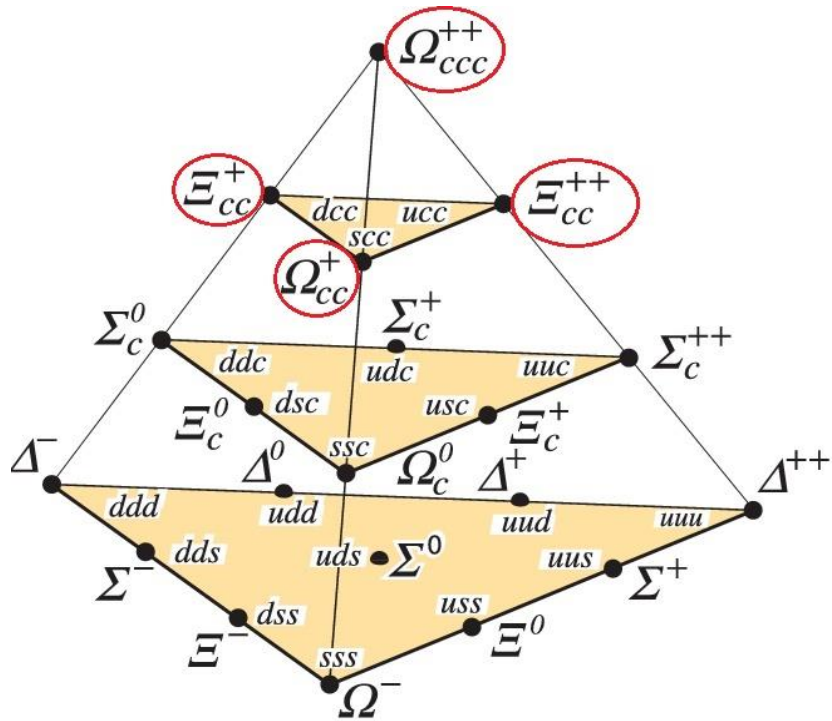

Fig. 1. *SU*(4) 20–plet of Baryons ($J^P=3/2^+$) made of *u*, *d*, *s* and *c* quarks, with an *SU*(3) de-couplet at the bottom [5].

### (A) MASS EQUAL SPACING RULE I

Gell-Mann developed this formula. Mass formula for de-couplets consists of equal spacing between the rows. The spaces are equal to an average value – 151MeV.

$$M_\Delta - M_{\Sigma^*} = M_{\Sigma^*} - M_{\Xi^*} = M_{\Xi^*} - M_\Omega \quad (3)$$

or

$$M_{\Sigma^*} - M_\Delta = M_{\Xi^*} - M_{\Sigma^*} = M_\Omega - M_{\Xi^*} = 151$$

or

$$\frac{M_{\Sigma^*} - M_\Delta}{M_{\Xi^*} - M_{\Sigma^*}} = \frac{M_\Omega - M_{\Xi^*}}{151\, MeV} \quad (4)$$

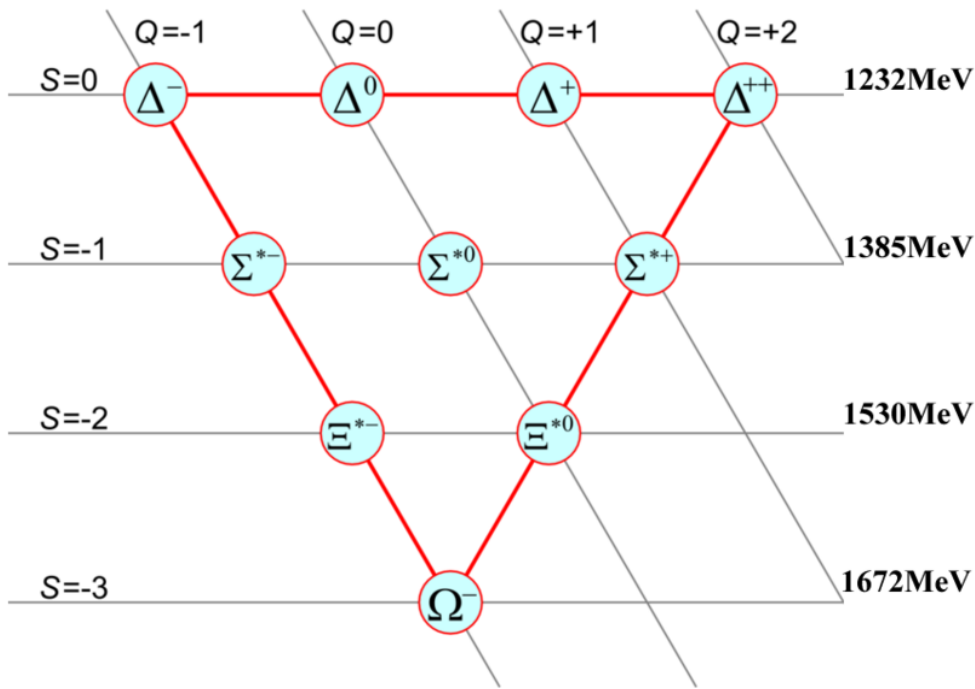

Fig. 2. $SU(3)$ de-couplet of Baryons ($J^P = 3/2^+$) made of $u$, $d$, and $s$ quarks, used by Gell Mann [5].

Gell-Mann used this formula and predicted the mass of the $\Omega^-$ baryon in 1962, equal to $M_\Omega = 1685 MeV$ [2]. Whereas actual mass of the $\Omega^-$ hyperon is equal to 1672 MeV, observed in 1964 [4]. Their mass difference is only 0.72 %, or in other words it was 99% true guess.

### (B) MASS EQUAL SPACING RULE II

I have developed another rule of equal spacing in layers of the particles with same charge number. According to the charge distribution, average masses of the charge layers are given by;

$$M_{-1} - M_0 = M_0 - M_{+1} = M_{+1} - M_{+2} = 74 \; MeV/c^2 \quad (5)$$

Here subscripts show the charge of the layers containing particles with same charge.

### (C) MASS EQUAL SPACING RULE III

In case of SU(4) spin half particles 20 plet, figure 2, we have;

Mass of $\Xi_{cc}^{++}$ (ucc) particle = 3621 MeV/c$^2$

Mass of $\Xi_c^+$ (usc) particle = 2468 MeV/c$^2$

Mass of $\Xi^0$ (uss) particle = 1315 MeV/c$^2$

Mass difference between $\Xi_{cc}^{++}$ and $\Xi_c^+$ states = 3621−2468 = 1153 MeV/c$^2$

And mass difference between $\Xi_c^+$ and $\Xi^0$ states = 2468−1315 = 1153 MeV/c$^2$

Mass difference between these particles is exactly same i.e. 1153 MeV/c$^2$. There difference in nature is only of one c-quark. So we can use such behavior for other multiplets' content particles too.

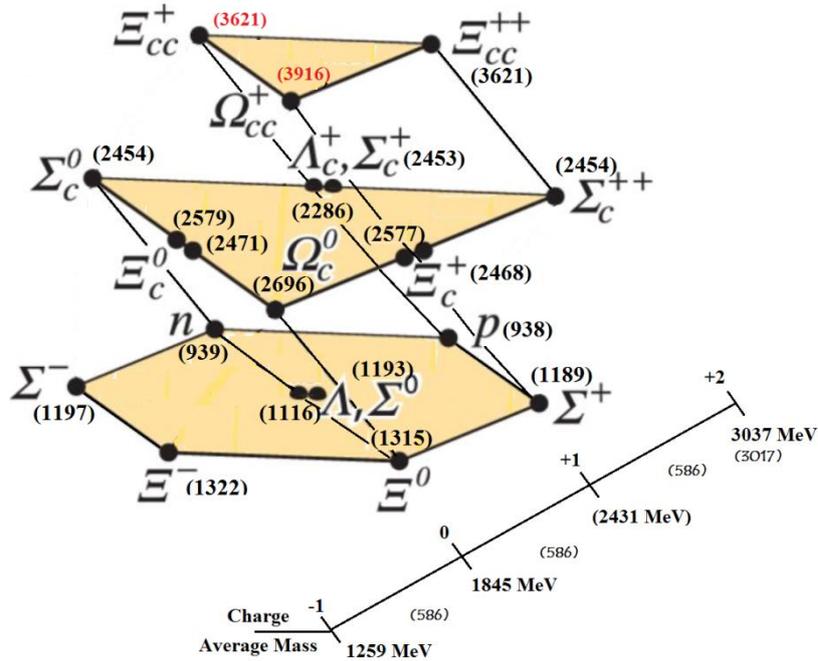

**Fig. 3.** *SU*(4) 20–plet of Baryons ($J^P$=1/2$^+$) made of *u*, *d*, *s* and *c* quarks, with different layers against charge and their average masses [7].

The twenty particles having spin 3/2 and even parity +1 forming one of the $SU(4)$ multiplets have been studied in reference [6] by using a lengthy complex method. Now we are interested to study the same twenty particles having spin 3/2 and even parity +1 using a very simple and interesting method to extract masses of the unknown particles predicted by standard model of particle physics. These particles are in their ground states, with $l = 0$. For simplicity, natural unit of the mass is used in this article at many places, that is MeV, instead of MeV/$c^2$.

Table 2: Mass difference between masses of hyperons having charmness 1 and zero.

| S.# | Particle with c | Particle with s | c Particle's mass (MeV/$c^2$) | s Particle's mass (MeV/$c^2$) | Mass Difference (MeV/$c^2$) |
|---|---|---|---|---|---|
| 1 | $\Sigma_c^{*++}$ (uuc) | $\Sigma^{*+}$ (uus) | 2518 | 1383 | 1135 |
| 2 | $\Xi_c^{*+}$ (usc) | $\Xi^{*0}$ (uss) | 2646 | 1532 | 1114 |
| 3 | $\Xi_c^{*0}$ (dsc) | $\Xi^{*-}$ (dss) | 2646 | 1535 | 1111 |
| 4 | $\Omega_c^{*0}$ (ssc) | $\Omega^{-}$ (sss) | 2766 | 1672 | 1094 |

Mass differences between masses of hyperons having charmness equal to 1 and zero are calculated using above stated technique and presented in table 2. Other quark contents of both compared particles are same. Only difference is of s-quark and c-quark. It is found that average mass difference is equal to 1114, ranging from 1094 to 1135 MeV/$c^2$. Actual mass difference between these quark masses is equal to 1270-93.4= 1176.6 MeV/$c^2$.

**Table 3:** Masses of missing hyperons having two charm quarks are calculated via adding mass differences into mass of particles with one charm quark.

| S.# | Particle with cc | Particle with c | Particle's mass Method | Particle's mass (MeV/$c^2$) |
|---|---|---|---|---|
| 1 | $\Xi_{cc}^{*++}$ (ucc) | $\Xi_c^{*+}$ (usc) | 2646+1111 | 3757±61 |
| 2 | $\Xi_{cc}^{*+}$ (dcc) | $\Xi_c^{*0}$ (dsc) | 2646+1114 | 3760±61 |
| 3 | $\Omega_{cc}^{*+}$ (scc) | $\Omega_c^{*0}$ (ssc) | 2766+1094<br>2766+1114 | 3870±62 |

Then masses of missing hyperons having two charm quarks are calculated via adding mass differences into mass of particles with one charm quark. These are presented in table 3. Thus we have obtained masses of three particles. One particle's $\Omega_{ccc}^{*++}$ mass is still not predicted. For this purpose we follow the following procedure.

In the *SU*(3) framework the Gell-Mann / Okubo relation for the $J^P = 1/2^+$ Octets, equation (1) and the equal spacing rule for the $J^P = 3/2^+$ de-couplets, equation (2) work so nicely that we cannot abandon linear mass formulae for baryons [7]. Same behavior of the mass splitting of the particles is also observed to get expression for the particles having $J^P = 1/2^+$ and forming the multiplet in SU(4) [7]. Similarly the mass splitting of the particles may also be used to get expression for the particles having $J^P = 3/2^+$ and forming the multiplet in SU(4) as shown in figure (1). Figure (1) can be viewed from another angle, as shown in figure (4). It is distributed in four different layers with increasing charge number.

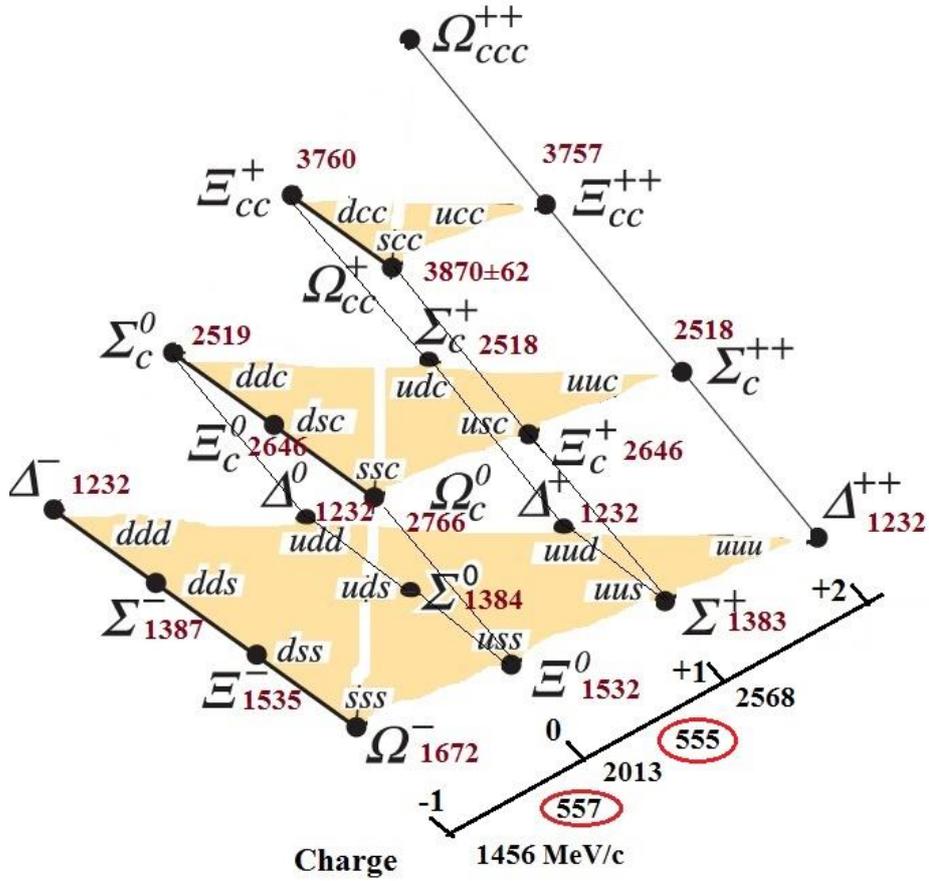

**Fig. 4.** *SU*(4) 20–plet of Baryons ($J^P=3/2^+$), with different layers against charge and their average masses.

## III    MASS OF THE $\Omega_{ccc}^{*++}$ HYPERON

In figure (4) quark contents of the particles along with masses of the observed particles and calculated masses are presented. Average masses of particles in different layers with same charge number is written in front of each layer at the bottom right. Since $\Omega_{ccc}^{++}$ is not discovered yet, therefore we cannot calculate average mass of the layer with charge number +2. However we can approximate masses of this unknown particle with simple method, similar to the method used by Gell-Mann/ Okubo for SU(3) de-couplet [8]. Average masses of the particles in layers with charge number -1 and 0, are given by $M_{-1}$ = 1456 MeV, $M_0$ = 2013 MeV and $M_{+1}$ = 2568 MeV respectively. Average mass of the particles in layer with charge number +2, $M_{+2}$ cannot be obtained due to mass of one missing particle.

Difference between average masses of the layers with charge number 0 and -1 is; $M_0 - M_{-1} = 557\ MeV$. Difference between average masses of the layers with charge number -1 and zero is; $M_{+1} - M_0 = 555\ MeV$. Adding this value into the average mass of the layer with charge number +1, i.e. $M_{+1}$ = 2568 MeV gives; $M_{+1} + 556\ MeV = 3124\ MeV = M_{+2}$. Hence we can say that there is an equal spacing rule between these layers, given by;

$$M_2 - M_1 = M_1 - M_0 = M_0 - M_{-1} = 556\ MeV \quad (6)$$

or

$$\boxed{M_n = M_{n-1} + 556\ MeV/c^2} \quad (7)$$

Where n= 0, 1, 2.

So average mass of the particles in layer with charge number +2, is obtained as equal to $M_{+2}$ = 3124 MeV.

Now total mass of the layer's particles should be:

$$Total\ Mass = 3124 \times 4 = 12496\ MeV/c^2$$

Mass of the particle is given by:

$$Mass\ of\ \Omega_{ccc}^{++} = 12496 - 3757 - 2518 - 1232 = 4989 MeV/c^2$$

$$Mass\ of\ \Omega_{ccc}^{++} = 4989 \pm 70 MeV/c^2$$

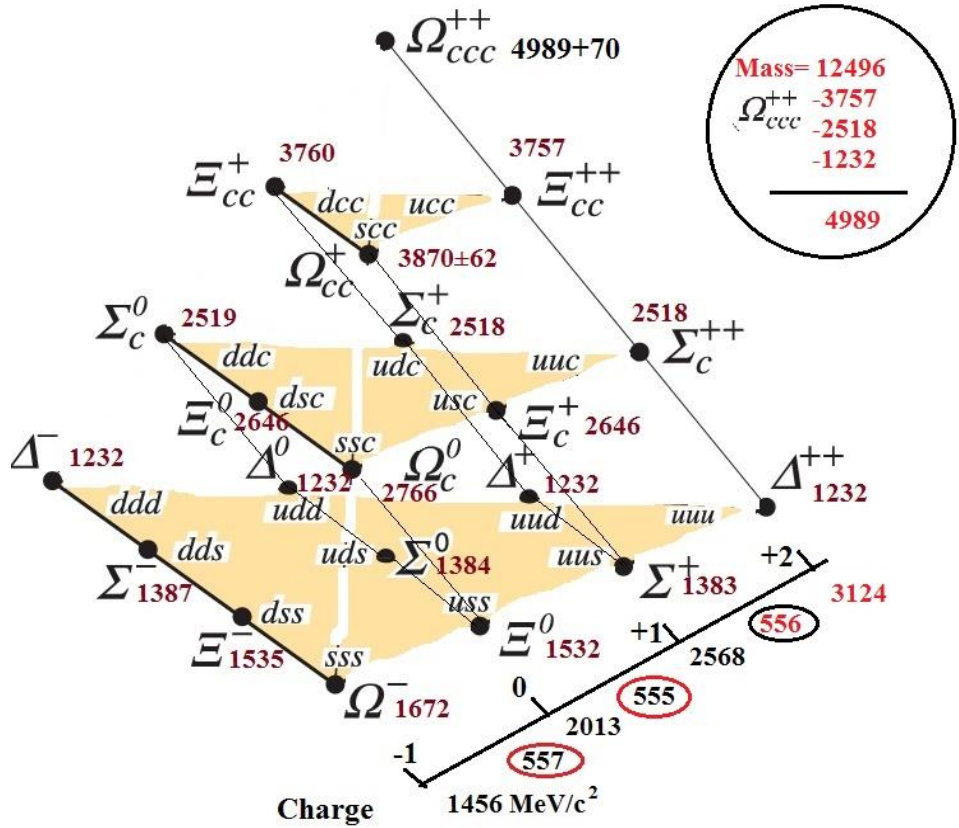

**Fig. 5.** $SU(4)$ 20–plet of Baryons ($J^P=3/2^+$) with particles masses.

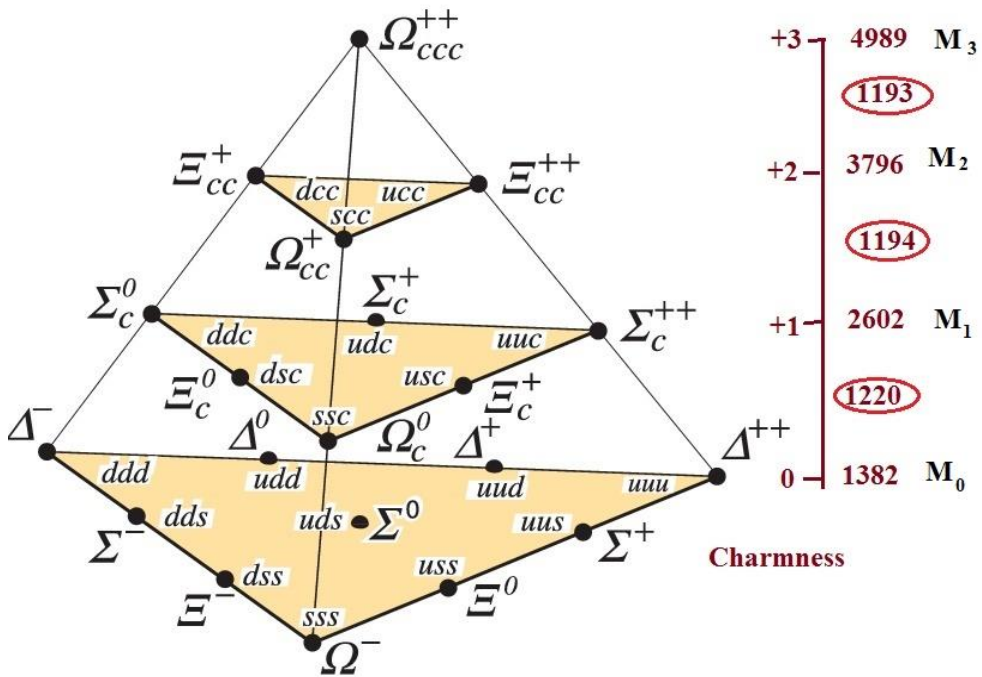

**Fig. 6.** SU(4) Multiplet, with increasing Mass, Charge number and Charmness from bottom to top.

## IV A GENERAL RELATION

The calculations may also be expanded to the whole 20- multiplet, as shown in Figs. 1 & 6, so that a combined formula for all the baryons masses of this multiplet may be extracted.

A general relation between layers of multiplet is given by;

$$M_3 - M_2 = M_2 - M_1 = M_1 - M_0 \approx 1200 \, MeV \quad (8)$$

or

$$\boxed{M_n = M_{n-1} + 1200 \, MeV/c^2} \quad (9)$$

Where n= 1, 2, 3.

And equation for masses of hadrons is given by;

$$\frac{\Xi_{cc}^* + \Xi^{*0,-}}{\Omega_{cc}^{*+} + \Sigma} = \frac{2(\Omega_c^{*0} + \Xi_c^{*0} + \Sigma_c + \Omega)}{3(\Delta + \Omega_{ccc}^{*++})} \approx 1$$

Thus the mass of $\Omega_{ccc}^{++}$ particle, estimated using the above method is equal to 4989±70 $MeV/c^2$. Statistical Errors are the square root values of the particle masses. Experimentally observed Particle masses may be slightly differ from the above estimated values of masses of all the four particles, due to rough estimates. However in that case the values obtained experimentally should agree with theoretical mass values with in two standard errors. All the predicted masses of the four unknown particles and their other properties are listed in Table 4.

**Table 4:** Approximated masses and other properties of the unknown particles in Standard Model of *SU*(4).

| S.# | Particles | Particle's mass ($MeV/c^2$) |
|---|---|---|
| 1 | $\Xi_{cc}^{*++}$ (*ucc*) | 3757±61 |
| 2 | $\Xi_{cc}^{*+}$ (*dcc*) | 3760±61 |
| 3 | $\Omega_{cc}^{*+}$ (*scc*) | 3870±62 |
| 4 | $\Omega_{ccc}^{*++}$ (*ccc*) | 4989±70 |

Now let us check what is the mass difference, $M_2 - M_1 = 1200 \, MeV/c^2$. It is actually mass of the Charm quark, which is the constituent of the particles. Charm number increases by +1 as we go up in the particles pyramid of Fig. 1, and Fig. 6. So this increase in the masses of the particles is due to the mass of the charm quark. Comparing this with the evaluated mass of the charm quark $M_c = 1270 \pm 30$ and $M_c = 1196 \pm 59 \pm 50 \; MeV/c^2$ listed in Ref. [5] gives satisfactory result.

## V  CONCLUSIONS

*SU*(4) multiplets of Standard Model have been analyzed in this article and some predictions have been made about the unknown masses and other properties of the $\Xi_{cc}^{*++}, \Xi_{cc}^{*+}, \Omega_{cc}^{*+}$, and $\Omega_{ccc}^{++}$ hyperons. Their masses are estimated to be $\Xi_{cc}^{*++}$ (3757 ± 61), $\Xi_{cc}^{*+}$ (3760 ± 61), $\Omega_{cc}^{*+}$ (3870 ± 62), and $\Omega_{ccc}^{++}$ (4989 ± 70) *MeV/c²*. Statistical errors are the square root values of the estimated masses of particles. Experimentally observed particles' masses may slightly differ from the estimated values of masses of all the four particles, due to rough estimates. However in that case the values obtained experimentally should agree with theoretical mass values with in two standard errors. A general formula for masses has been derived for the *SU*(4) multiplets, which gives satisfactory results for observed and estimated masses of the particles of *SU*(4). The mass difference in the multiplets of the *SU*(4) is due to the mass of Charm quark, and its value is also equal to mass of the charm quark. The results in this research, if proved correct, will be helpful in understanding the particle physics, and will strengthen the Standard Model of particles.